\documentclass[twocolumn,showpacs,prl]{revtex4}
\usepackage{mathrsfs}
\usepackage{graphicx,,booktabs}
\usepackage{color}
\usepackage{amssymb}

\begin{document}
\title{A novel explanation of charmonium-like structure in $e^+e^-\to \psi(2S)\pi^+\pi^-$}
\author{Dian-Yong Chen$^{1,2}$}
\author{Jun He$^{1,2}$}
\author{Xiang Liu$^{1,3}$\footnote{Corresponding author}}\email{xiangliu@lzu.edu.cn}
\affiliation{
$^1$Research Center for Hadron and CSR Physics,
Lanzhou University and Institute of Modern Physics of CAS, Lanzhou 730000, China\\
$^2$Nuclear Theory Group, Institute of Modern Physics of CAS, Lanzhou 730000, China\\
$^3$School of Physical Science and Technology, Lanzhou University, Lanzhou 730000,  China}
\date{\today}

\begin{abstract}

We first present a non-resonant description to charmonium-like
structure $Y(4360)$ in the $\psi(2S)\pi^+\pi^-$ invariant mass spectrum
of the $e^+e^-\to \psi(2S)\pi^+\pi^-$ process. The $Y(4360)$ structure
is depicted well by the interference effect of the production
amplitudes of $e^{+} e^{-} \to \psi(2S) \pi^+ \pi^-$ via the intermediate
charmonia $\psi(4160)/\psi(4415)$ and direct $e^+e^-$ annihilation
into $\psi(2S)\pi^+ \pi^-$. This fact shows that $Y(4360)$ is not a
genuine resonance, which naturally explains why $Y(4360)$ was only
reported in its hidden-charm decay channel $\psi(2S)\pi^+\pi^-$ and
was not observed in the exclusive open-charm decay channel or $R$-value
scan.

\end{abstract}


\maketitle
With the observations of charmonium-like states $X$, $Y$, $Z$ by
experiment, studying charmonium-like states $X$, $Y$, $Z$ has became
an intriguing and important research topic in hadron physics at
present. The initial-state radiation (ISR) processes $e^+e^-\to
\gamma_{_{ISR}} J/\psi \pi^+\pi^-$ and $e^+e^-\to \gamma_{_{ISR}}
\psi(2S) \pi^+\pi^-$ are the ideal platform to carry out the study
of charmonium-like states with $J^{PC}=1^{--}$. Besides the first
observation of the $Y(4260)$ structure in $e^+e^-\to \gamma_{_{ISR}}
J/\psi \pi^+\pi^-$ announced by BaBar \cite{Aubert:2005rm} and
confirmed by the CLEO and Belle Collaborations
\cite{Coan:2006rv,He:2006kg,:2007sj}, a new charmonium-like
structure $Y(4360)$ was reported at BaBar \cite{Aubert:2006ge} by
analyzing the $\psi(2S) \pi^+\pi^-$ invariant mass spectrum
of $e^+e^-\to \gamma_{_{ISR}} \psi(2S) \pi^+\pi^-$. Later, Belle
confirmed BaBar's result and also announced another structure
$Y(4660)$ in $e^+e^-\to \gamma_{_{ISR}} \psi(2S) \pi^+\pi^-$
\cite{:2007ea}.

These observations of $Y(4360)$ and $Y(4660)$ not only make the
vector charmonium-like structures in ISR processes become abundant,
but also stimulate theorist's interest in revealing their underlying
structure. Ding {\it et al.} indicated that $Y(4360)$ could be as a
$n^{2S+1}J_L=3^3D_1$ charmonium state or a charmonium hybrid while
$Y(4660)$ is a good candidate of charmonium with
$n^{2S+1}J_L=5^3S_1$ \cite{Ding:2007rg}. With these possible
structure assignments, the corresponding decay behaviors of
$Y(4360)$ and $Y(4660)$ are given \cite{Ding:2007rg}. Under the same
charmonium assignments to $Y(4360)$ and $Y(4660)$ as that in Ref.
\cite{Ding:2007rg}, the di-electron widths of $Y(4360)$ and
$Y(4660)$ are obtained \cite{Badalian:2008dv}. With
screened potential, Li and Chao calculated the mass spectrum of
charmonium, which explained $Y(4360)$ and $Y(4660)$ as $\psi(3D)$
and $\psi(6S)$ charmonia respectively and predicted their decay
behaviors \cite{Li:2009zu}. In Ref. \cite{Segovia:2008zz}, $Y(4360)$ was also explained as a $4S$ charmonium state. The baryonium
assignments to $Y(4360)$ and $Y(4660)$ were proposed in
\cite{Qiao:2007ce}, where $Y(4360)$ might be the radial excited
state of $Y(4260)$ and $Y(4660)$ is the radial excited state of the
C-spin singlet with the flavor wave function
$(|\Lambda_c^+\Lambda_c^-\rangle+|\Sigma_c^0\bar{\Sigma}_{c}^0\rangle)/\sqrt{2}$,
which in fact are a extension of former $\Lambda_c^{+}
\Lambda_{c}^{-}$ baryonium explanation to $Y(4260)$
\cite{Qiao:2005av}. Associated with the charmonium-like state
$Y(4630)$ in $e^+e^-\to \Lambda_c^+{\Lambda}_c^-$
\cite{Pakhlova:2008vn} with $Y(4660)$, Cotugno {\it et al.}
suggested that both $Y(4630)$ and $Y(4660)$ are from the same
charmed baryonium state \cite{Cotugno:2009ys}. The authors in Ref.
\cite{Kalashnikova:2008qr} indicated that the vector hybrid
charmonium with strong coupling with $D^*D_0$ channel results in
considerable threshold attraction related to $Y(4360)$. By the
QCD sum rule, Albuquerque and Nielsen calculated the masses of
$(c\bar{c}q\bar{q})$ or $(c\bar{c}s\bar{s})$ states with $J^{PC}
=1^{--}$, and found that $(c\bar{c}s\bar{s})$ state with obtained
mass $4.65$ GeV corresponds to $Y(4660)$ while $Y(4360)$ could be as a
$D_0\bar{D}^{*0}$ molecular state \cite{Albuquerque:2008up}. Later,
the mass of the P-wave $cs$-scalar-diquark
$\bar{c}\bar{s}$-scalar-antidiquark state was computed in the
framework of the QCD sum rule. The result $4.64 \pm 0.25$ GeV is in good
agreement with the experimental value of $Y(4660)$
\cite{Zhang:2010mw}. With the obtained effective potential of
$D^*\bar{D}_1$ system in Refs. \cite{Liu:2007bf,Liu:2008xz}, Close
{\it et al.} explained $Y(4360)$ as
a $2S$ $D^*\bar{D}_1$ molecular state \cite{Close:2009ag,Close:2010wq}. $Y(4660)$ as an $f_0(980)\psi(2S)$ bound state was proposed \cite{Guo:2008zg,Guo:2010tk} and studied \cite{Wang:2009hi}.

Although different explanations for $Y(4360)$ and $Y(4660)$ were
given from the point of view of conventional charmonium or exotic
state just summarized in the above, the current situation regarding
the $1^{--}$ charmonium-like structures $Y(4360)$ and $Y(4660)$
produced through ISR process is clearly unsettled
\cite{Albuquerque:2008up}. No matter what explanations we proposed,
we can not avoid the fact that there is not any evidence of these
$1^{--}$ charmonium-like structures produced through ISR process in the
exclusive open-charm decay channel
\cite{Abe:2006fj,Pakhlova:2008zza,Pakhlova:2007fq,:2009jv,Aubert:2006mi,:2009xs,CroninHennessy:2008yi}
and $R$-value scan \cite{Ablikim:2007gd}.

Recently, we proposed a new approach different from conventional
charmonium and exotic state explanations to explain $Y(4260)$
structure, where $Y(4260)$ is not a genuine resonance
\cite{Chen:2010nv}. $e^+e^-\to J/\psi\pi^+\pi^-$ can occur directly
via $e^+e^-$ annihilation. In addition, the intermediate charmonia
can contribute to $e^+e^-\to J/\psi\pi^+\pi^-$, where we choose
$\psi(4160)$ and $\psi(4415)$ as intermediate charmonia since
the $Y(4260)$ structure is just sandwiched by two known charmonia
$\psi(4160)$ and $\psi(4415)$. The interference effect from the
above two production mechanisms for $e^+e^-\to J/\psi\pi^+\pi^-$ can
reproduce the $Y(4260)$ structure in the $J/\psi\pi^+\pi^-$ invariant
mass spectrum of the $e^+e^-\to J/\psi\pi^+\pi^-$ process well
\cite{Chen:2010nv}. Such non-resonant explanation can naturally
answer why experiment only reported $Y(4260)$ in its hidden-charm
decay channel.

Successfully explaining $Y(4260)$ structure stimulates our interest
in extending the same picture to describe the charmonium-like
structures observed in $e^+e^-\to \gamma_{_{ISR}} \psi(2S)
\pi^+\pi^-$ since the experimental situation of $Y(4360)$ is similar
to that of $Y(4260)$. If the $Y(4360)$ structure also be depicted by the
non-resonant explanation similar to that proposed in our former work
\cite{Chen:2010nv}, it not only further supports non-resonant
explanation to $Y(4260)$ but also deepens our understanding of the
underlying mechanism resulting in these charmonium-like structures
produced via ISR process.

\begin{figure}[htb]
\centering \scalebox{0.5}{\includegraphics{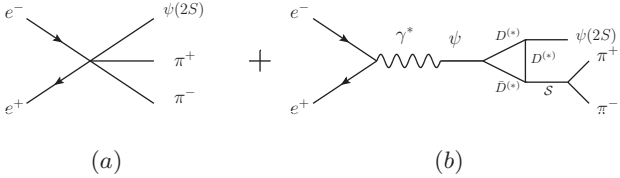}}
\put(-200,-20){$(a)$}%
\put(-70,-20){$(b)$}%
\caption{The diagrams depicting the $e^{+} e^{-} \to
\psi(2S) \pi^+ \pi^-$ production process. (a) The $e^+e^-$
annihilation directly into $\psi(2S)\pi^+\pi^-$. (b) The
contributions from intermediate charmonia to $e^{+} e^{-} \to
\psi(2S) \pi^+ \pi^-$. Here, the hadronic loop effect plays
important role to the $\psi\to \psi(2S)\mathcal{S}$ transition.
\label{Fig-Feyn1}}
\end{figure}

In general, the $e^+e^-\to \psi(2S)\pi^+\pi^-$ process occurs via two
mechanisms. One is the direct production of $\psi(2S)\pi^+\pi^-$ by
$e^+e^-$ annihilation (Fig. \ref{Fig-Feyn1}(a)),
which can be depicted by the production amplitude
\begin{eqnarray}
\mathscr{A}_{{NoR}} = g_{{NoR}} \bar{u}(-k_1) \gamma_{\mu} u(k_2)
\epsilon_{\psi(2S)}^{\mu} (k_5)\, \mathcal{F}_{{NoR}}(s),\label{1}
\end{eqnarray}
where the form factor $\mathcal{F}_{{NoR}}(s)$ is introduced
to represent the $s$-dependence of $\psi(2S) \pi^+ \pi^-$ production directly via the
$e^+e^-$ annihilation, which can be represented as
$\mathcal{F}_{{NoR}} (s) =\mathrm{exp} \left(-a
(\sqrt{s}-\sum_f m_f)^2 \right)
$
with $\sum_f m_f$ as the sum of the masses of the final states for
$e^+e^-\to \psi(2S)\pi^+\pi^-$. In Eq. (\ref{1}), two
parameters $a$ and coupling constant $g_{\mathrm{NoR}}$ are introduced. $\sqrt{s}$
is the energy of center of mass frame of $e^+e^-$. $k_1,\,k_2,\,k_5$
correspond to the four momenta of $e^+,\,e^-,\,\psi(2S)$,
respectively.

Another is from intermediate charmonium contribution shown in
Fig. \ref{Fig-Feyn1}(b), where $e^+$ and $e^-$ annihilate into
one virtual photon, which interacts with vector charmonium. Then,
charmonium transits into $\psi(2S)\pi^+\pi^-$. Since two well-known
charmonia $\psi(4160)$ and $\psi(4415)$ are most close to $Y(4360)$
among these observed charmonia, we choose $\psi(4160)$ and
$\psi(4415)$ as the intermediate states to $e^+e^-\to
\psi(2S)\pi^+\pi^-$ process. Under the vector meson dominance (VMD)
mechanism \cite{Bauer:1977iq,Bauer:1975bw} for the $\gamma\to
\psi(4160)/\psi(4415)$ coupling, the general amplitude of
$e^+(k_1)e^-(k_2)\to \psi(2S)(k_5)\pi^+(k_3)\pi^-(k_4)$ via the
intermediate charmonia reads as
\begin{eqnarray}
\mathscr{A}_{\psi,\mathcal{S}} &=& \bar{u}(-k_1) e \gamma^{\mu}
u(k_2)  \frac{e\,
m_\psi^2/f_{\psi}\,\epsilon_{\psi(2S)}^{\rho}}{(k_1+k_2)^2-m_{\psi}^2
+ i
m_{\psi} \Gamma_{\psi}}  \nonumber\\
&&\times  \Big[g^\psi_{A} g_{\nu \rho} k_{5}
\cdot (k_3+k_4) +g_{B}^\psi k_{5 \nu} (k_{3 \rho} +k_{4 \rho}) \Big]
\nonumber\\
&&\times \frac{-g_{\mu}^{\,\,\nu}}{ (k_1 +k_2)^2}\frac{ g_{\mathcal{S} \pi \pi} (k_3
\cdot k_4)}{(k_3+k_4)^2-m_{\mathcal{S}}^2 +
im_{\mathcal{S}} \Gamma_{\mathcal{S}}} \label{dd}
\end{eqnarray}
with the decay constant $f_{\psi}$ of intermediate charmonium and
the coupling constant $g_{\mathcal{S} \pi\pi}$ of scalar state interacting with dipion, where we assume $\pi^+\pi^-$ from scalar
$\sigma$ and $f_0(980)$ states which is consistent with the measurement result of the
$\pi^+\pi^-$ invariant mass spectrum in $e^+e^-\to
\psi(2S)\pi^+\pi^-$ \cite{:2007ea}. Thus, in Eq. (\ref{dd}) we set
$\psi =\{\psi_1=\psi(4160), \psi_2=\psi(4415)\}$ and
$\mathcal{S}=\{\sigma, f_{0}(980)\}$. In Eq. (\ref{dd}), the coupling
constants $g^\psi_{A}$ and $g^\psi_{B}$ are relevant to the
transition of intermediate charmonium coupling to
$\psi(2S)\mathcal{S}$, which are determined by evaluating hadronic
loop mechanism
\cite{Liu:2006dq,Liu:2009dr,Liu:2006df,Liu:2008yy,Liu:2009iw}. Here,
$g^\psi_{A}$ and $g^\psi_{B}$ are the functions of $\beta_1$ and
$\beta_2$ respectively, where we adopt $\beta_1=\beta_2=1$
\cite{Cheng:2004ru}, which are the parameters from the introduced
form factor in hadronic loop calculation (see Ref.
\cite{Chen:2010nv} for more details and the input parameters)

We write out the total amplitudes for $e^{+} e^{-} \to \psi(2S) \pi^+ \pi^-$
\begin{eqnarray}
\mathscr{A}_{tot} &=& \mathscr{A}_{NoR}+ e^{i \phi_1}
\Big(\mathscr{A}_{\psi_1, \sigma} + e^{i \phi_{s}} \mathscr{A}
_{\psi_1, f_0} \Big) \nonumber\\
&&\hspace{0mm}+ e^{i \phi_2} \Big(\mathscr{A} _{\psi_2, \sigma} +
e^{i \phi_{s}} \mathscr{A} _{\psi_2, f_0} \Big)\nonumber\\
&\equiv&\mathscr{A}_{NoR}+\mathscr{A}_{\psi_1}+\mathscr{A}_{\psi_2}\label{amp}
\end{eqnarray}
with three introduced phase angles $\phi_1$, $\phi_2$, $\phi_{s}$.
Generally speaking the phases between different Feynman diagrams
are fixed and not arbitrary as they are in this work. However, there
maybe exist hadronic loop effects which generate different phases among the
different diagrams (because the momentum flow is different). Hence it is
permissible to parameterize our ignorance of these interactions with
these arbitrary phases to be fitted as is done in this work \cite{Chen:2010nv}. The observable is proportional to the modulus square of production amplitude $\mathscr{A}_{tot}$, which includes
the squared amplitudes ($|\mathscr{A}_{NoR}|^2$, $|\mathscr{A}_{\psi_1}|^2$, $|\mathscr{A}_{\psi_2}|^2$)
and the cross terms ($2\mathrm{Re}(\mathscr{A}_{\psi_1}\mathscr{A}_{NoR}^*)$, $2\mathrm{Re}(\mathscr{A}_{\psi_2}\mathscr{A}_{NoR}^*)$, $2\mathrm{Re}(\mathscr{A}_{\psi_2}\mathscr{A}_{\psi_1}^*)$), where such cross terms reflect the interference effect just mentioned above.

With the above preparation, in the following we
investigate whether the experimental data given by BaBar and Belle
\cite{Aubert:2006ge,:2007ea} can be recovered by our model, where
five parameters listed in Table \ref{chi} are applied to fit the
experimental data of $e^{+} e^{-} \to \psi(2S) \pi^+ \pi^-$ at BaBar
and Belle.

\begin{figure*}[hbtp]
\begin{center}
\begin{tabular}{c}
\includegraphics[height=300pt]{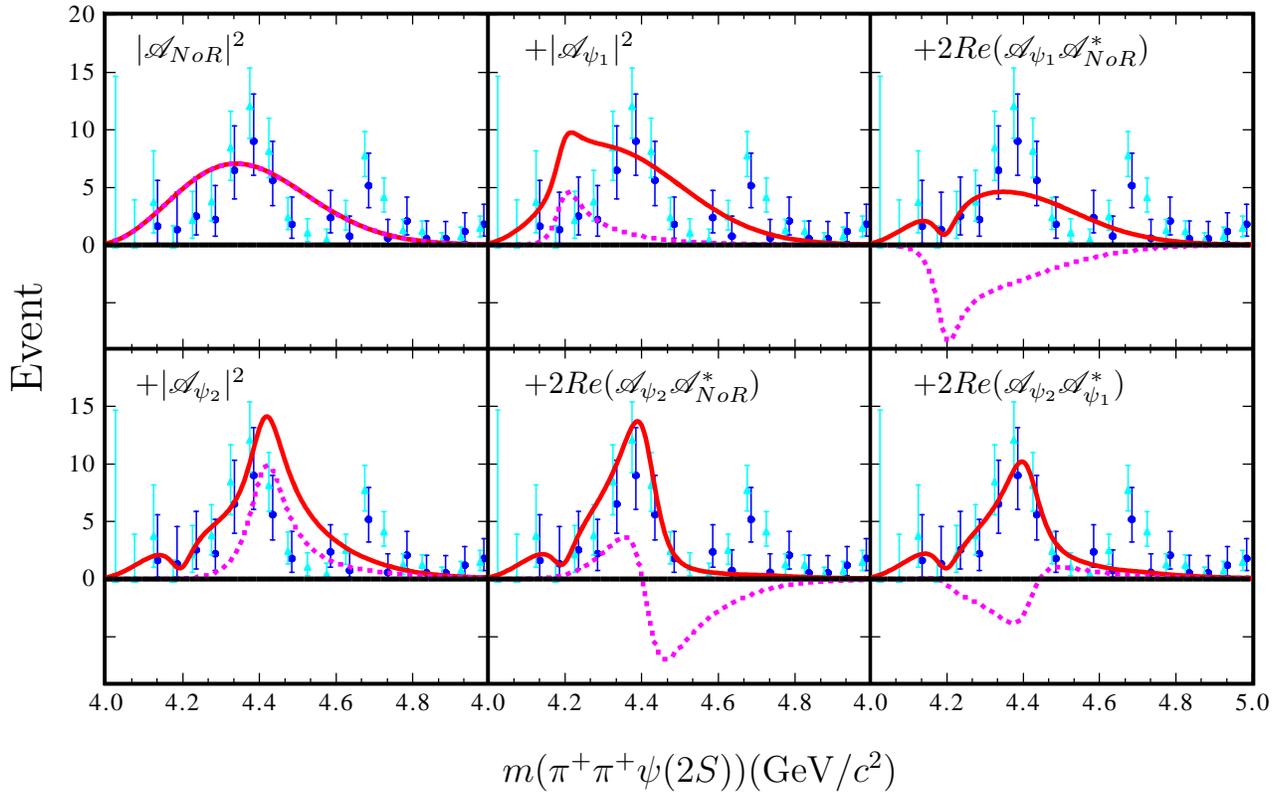}\\ [-20pt]
\end{tabular}
\end{center}
\caption{(Color online.) The line shape of the cross section of
$e^+e^-\to \psi(2S) \pi^+\pi^-$ process dependent on
$\sqrt{s}=m(\pi^+\pi^-\psi(2S))$. The blue and cyan points with
error bar are experimental data measured by BaBar and Belle, respectively. Our
result is normalized to the experimental data. Here, one also
presents the changes of the theoretical line shape (red solid line) by
adding the contributions from these six terms
($|\mathscr{A}_{NoR}|^2$, $|\mathscr{A}_{\psi_1}|^2$,
$2\mathrm{Re}(\mathscr{A}_{\psi_1}\mathscr{A}_{NoR}^*)$,
$|\mathscr{A}_{\psi_2}|^2$,
$2\mathrm{Re}(\mathscr{A}_{\psi_2}\mathscr{A}_{NoR}^*)$ and
$2\mathrm{Re}(\mathscr{A}_{\psi_2}\mathscr{A}_{\psi_1}^*)$) one by
one. The sum of such six terms finally results in the total line
shape of the cross section of $e^+e^-\to J/\psi\pi^+\pi^-$ process
just shown in the last diagram. The pink dashed line in each diagram
reflects the contribution of each of six terms in
$|\mathscr{A}_{tot}|^2$. \label{Fig-Fita}} \label{x4260}
\end{figure*}
\refstepcounter{figure}

As shown in the last diagram of Fig. \ref{Fig-Fita}, we obtain the best fit (red solid line)
for the experimental data of $e^{+} e^{-} \to \psi(2S) \pi^+ \pi^-$, where
the obtained central values with errors of the fitting parameters are listed in Table \ref{chi},
which indicate that the theoretical line shape of $e^{+} e^{-} \to \psi(2S) \pi^+ \pi^-$ is not sensitive to
the changes of the fitting parameters. Thus, our theoretical study indicates that the $Y(4360)$
structure in the $\psi(2S)\pi^+\pi^-$ invariant mass spectrum is
reproduced well under the interference effect proposed in this work.

For explicitly illustrating the evolution of, we present how the
$Y(4360)$ signal in $e^{+} e^{-} \to \psi(2S) \pi^+ \pi^-$ is
depicted by adding the intermediate charmonia $\psi(4160)/\psi(4415)$ and
considering the interference effect in a stepwise fashion, which is given by the remaining five diagrams in Fig. \ref{Fig-Fita}. Here, the obtained fitting parameters (see Table \ref{chi}) corresponding to the best fit are adopted.
We notice that the interference
of production amplitudes of the $e^+e^-\to \psi(2S)\pi^+\pi^-$
processes via direct $e^+e^-$ annihilation and through the intermediate
charmonia $\psi(4160)/\psi(4415)$ plays a crucial role if fitting
experimental line shape of the $Y(4360)$ structure. Describing the
$Y(4360)$ structure well in our model shows that the $Y(4360)$
structure in $e^+e^-\to \psi(2S) \pi^+\pi^-$ is not genuine
resonance, which naturally explains why there is no any evidence of
$Y(4360)$ in the exclusive open-charm process and $R$-value scan.
As announced by Belle, there is another structure $Y(4660)$ besides $Y(4360)$ in the $\psi(2S)\pi^+\pi^-$ invariant mass spectrum. Our study shows that the $Y(4660)$ structure can not described by the interference effect just mentioned in this work. Revealing the property of $Y(4660)$ is still an intriguing research topic.

\begin{table}[htb]
\centering%
\caption{The values of parameters for the best fit (the red solid
line shape in Fig. \ref{Fig-Fita}) to the experimental data. Here, GeV$^{-2}$ and GeV$^{-1}$ are as units of $a$ and $g_{NoR}$ respectively.
\label{chi}}
\begin{tabular}{cccccccc}
\toprule[1pt]
Parameter & Value &Parameter & Value (Rad)\\
\midrule[1pt]
$a$       &      $4.9248 \pm 0.4105$ & $\phi_1$  &    $2.6770 \pm 0.1260$ \\
$g_{NoR}$ &      $0.0074 \pm 0.0009$ & $\phi_2$  &    $1.8509 \pm 0.3010$ \\
          &                                     & $\phi_s$  &    $0.0003 \pm 0.2467$ \\
\bottomrule[1pt]
\end{tabular}
\end{table}

The whole picture of the $Y(4360)$ structure proposed in this letter
is an important extension of that for the $Y(4260)$ structure in
$e^+e^-\to J/\psi \pi^+\pi^-$, where $Y(4260)$ is also not genuine
resonance \cite{Chen:2010nv}. If comparing the experimental
information of $Y(4360)$ and $Y(4260)$, one notices that both
$Y(4360)$ and $Y(4260)$ were observed in exclusive hidden-charm
decay channel by ISR process, where $Y(4360)$ and $Y(4260)$
correspond to $\psi(2S)\pi^+\pi^-$ and $J/\psi\pi^+\pi^-$ channels
respectively, which reflects the small difference between $Y(4360)$
and $Y(4260)$. To some extent, such non-resonant picture for
$Y(4360)$ not only embodies the similarity between $Y(4360)$ and $Y(4260)$,
but also supports $Y(4260)$ non-resonant explanation in Ref.
\cite{Chen:2010nv}.

In summary, stimulated by the observation of charmonium-like
structure $Y(4360)$ in the $e^{+} e^{-} \to \psi(2S) \pi^+ \pi^-$
process, in this letter we first propose a novel non-resonant explanation
to the underlying structure of $Y(4360)$, which is different from
the previous conventional charmonium or exotic explanations. The
$Y(4360)$ structure is obtained by the interference effect of the
production amplitude of $e^{+} e^{-} \to \psi(2S) \pi^+ \pi^-$ via
the intermediate charmonia $\psi(4160)/\psi(4415)$ and direct $e^+e^-$
annihilation into $\psi(2S)\pi^+ \pi^-$. Such picture provides a
natural explanation of why experiment did not find any evidence of
$Y(4360)$ in the exclusive open-charm decay process and $R$-value scan.
Furthermore, the study presented in this letter further deeps our
understanding of charmonium-like structures observed in hidden-charm
process.

This study can be extended to include the theoretical study of other
charmonium-like structures appearing in other hidden-charm processes
from $e^+e^-$ annihilation by ISR mechanism. Recently, the CLEO-c
Collaboration announced a preliminary result by studying $e^+e^-\to
h_c\pi^+\pi^-$, where there is a suggestive rise at 4260 MeV in the
$h_c\pi^+\pi^-$ invariant mass spectrum \cite{Mitchell:2011zk}. Thus, we expect
more experimental measurements especially to $e^+e^-\to
h_c\pi^+\pi^-$. In addition, carrying out the study of the remaining
charmonium-like structure $Y(4660)$ \cite{:2007ea} in the
$\psi(2S)\pi^+\pi^-$ invariant mass spectrum will be also an
intriguing and valuable research topic by associating with
charmonium-like structure $Y(4630)$ in $e^+e^-\to
\Lambda_c^+{\Lambda}_c^-$ \cite{Pakhlova:2008vn}.

This project is supported by the National Natural Science Foundation of
China under Grants No. 10705001, No. 10905077, No. 11005129, No.
11035006, No. 11047606 and the Ministry of Education of China
(FANEDD under Grant No. 200924, DPFIHE under Grant No.
20090211120029, NCET under Grant No. NCET-10-0442, and the Fundamental
Research Funds for the Central Universities).

\end{document}